\documentclass[11pt]{article}
\usepackage{cite}
\usepackage{mdframed}
\usepackage{epsfig,subfigure}
\usepackage{amssymb}
\usepackage[fleqn]{amsmath}
\usepackage{color}
\usepackage{arydshln}
\def\QED{\mbox{\rule[0pt]{1.5ex}{1.5ex}}}

\usepackage{graphicx}
\usepackage{color}
\usepackage[dvipsnames]{xcolor}

\definecolor{armygreen}{rgb}{0.29, 0.33, 0.13}

\usepackage{amssymb}
\usepackage{amsmath,amsfonts,amssymb}
\usepackage{verbatim}
\usepackage{stfloats}
\usepackage[bookmarks=false]{}
\usepackage{algorithm}
\usepackage{algcompatible}

\newtheorem{theorem}{Theorem}

\newtheorem{lemma}{Lemma}

\newcommand\blfootnote[1]{%
  \begingroup
  \renewcommand\thefootnote{}\footnote{#1}%
  \addtocounter{footnote}{-1}%
  \endgroup
}

\begin{document}
\date{}

\title{
The Capacity of Anonymous Communications
}
\author{\normalsize Hua Sun \\
}

\maketitle

\blfootnote{This paper will be presented in part at ISIT 2018. Hua Sun (email: hua.sun@unt.edu) is with the Department of Electrical Engineering at the University of North Texas. }

\maketitle

\begin{abstract}
We consider the communication scenario where $K$ transmitters are each connected to a common receiver with an orthogonal noiseless link. One of the transmitters has a message for the receiver, who is prohibited from learning anything in the information theoretic sense about which transmitter sends the message 
(transmitter anonymity is guaranteed).  
The capacity of anonymous communications is the maximum number of bits of desired information that can be anonymously communicated per bit of total communication. For this anonymous communication problem over a parallel channel with $K$ transmitters and 1 receiver, we show that the capacity is $1/K$, i.e., to communicate 1 bit anonymously, each transmitter must send a 1 bit signal. Further, it is required that each transmitter has {\color{black}at least} 1 bit correlated randomness (that is independent of the messages) per message bit and the size of correlated randomness at all $K$ transmitters is at least $K-1$ bits per message bit. 
\end{abstract}

\newpage

\allowdisplaybreaks
\section{Introduction}
Traditional studies in information theoretic security and cryptography focus on efficient coding techniques for protecting the information contents. There is much recent interest in shifting the objective to hide user behaviors. For example, private information retrieval (PIR) aims to pursue communication efficient methods for hiding the identity of the desired message that the user wants to retrieve from a set of distributed replicated databases. The fundamental capacity limits of PIR and several of its variants are characterized recently in \cite{Sun_Jafar_PIR, Sun_Jafar_TPIR, Banawan_Ulukus}.

In this work, we consider the anonymous communication problem, where the goal is to hide the identity of the transmitters, receivers and the association between the two in a network. This problem of anonymous communications has been studied extensively in cryptography and computer science communities \cite{Peng, Ren_Wu, Danezis_Diaz}, where typically the objective is to provide scalable solutions over large networks while information theoretic optimality guarantees are not considered or treated in the approximate order sense. 

\begin{figure}[h]
\begin{center}
\includegraphics[width=3 in]{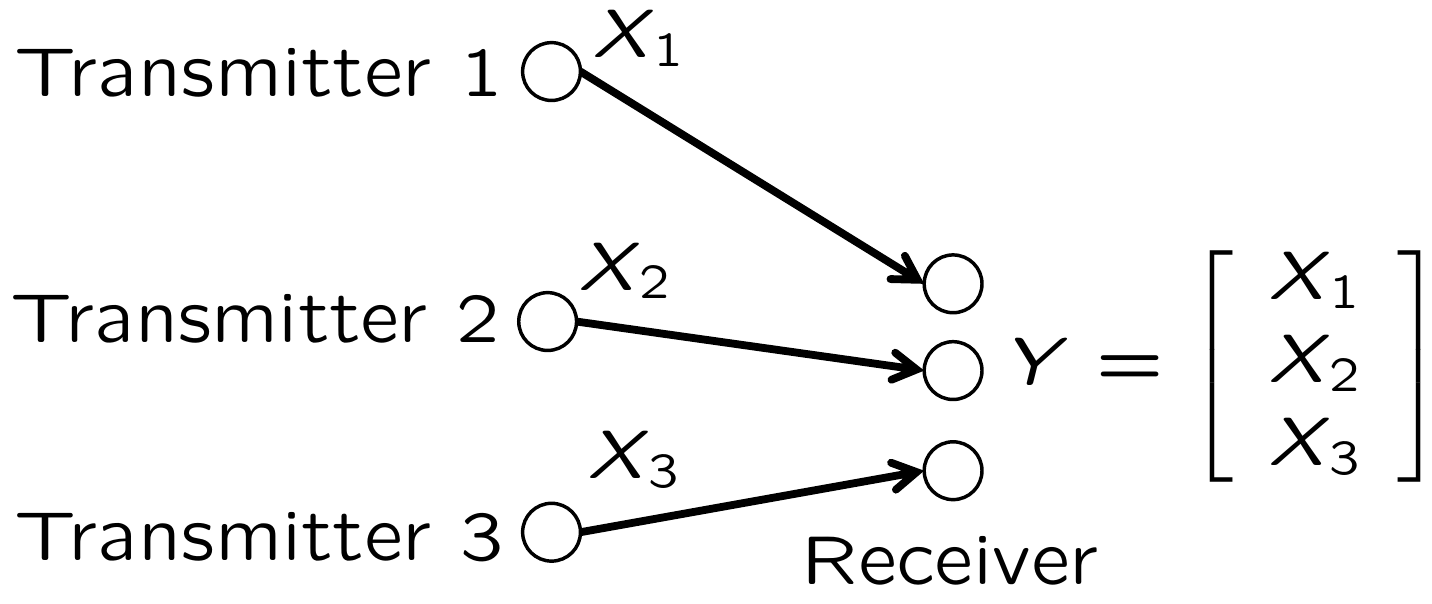}
\caption{\small Network topology: transmitters are connected to a single receiver with parallel interference-free noiseless links.}
\label{fig:model}
\end{center}
\end{figure}

We focus on an elemental model where $K$ transmitters want to communicate to a common receiver anonymously with interference-free noiseless parallel channels\footnote{Separate and perfect communication links are the least favorable channel conditions for anonymity because this assumption eliminates the possibility of hiding over direct interactions between the signals and noise.}. Our goal is to identify the exact information theoretic limits on the rate and common randomness for anonymous communications.
For example, consider the case where we have $K = 3$ transmitters. As each transmitter is connected to the receiver with a parallel channel, the received signal $Y$ is the collection of all transmitted signals, $X_1, X_2, X_3$ (see Figure \ref{fig:model}). 

\bigskip

One of the transmitters wishes to send a desired message to the receiver without being identified, i.e., the receiver decodes the message correctly, but has no knowledge about which transmitter sends the message. This anonymity constraint requires that no matter which transmitter wants to send the message, the received signal must be identically distributed and the decoding mapping can not depend on the desired transmitter index. To accomplish the task of keeping the transmitter identity anonymous, we assume that the transmitters share some correlated random variables that are independent of the messages. In this case, we assume that Transmitter 1 holds $a$, Transmitter 2 holds $b$ and Transmitter 3 holds $a+b$, where $a, b$ are two i.i.d uniform random bits (that form the correlated random variables). Then a simple scalar linear coding scheme that guarantees transmitter anonymity is presented next. Suppose the desired transmitter index is $\theta \in \{1,2,3\}$. The transmitted signals are
\begin{eqnarray}
X_1 &=& a + 1(\theta = 1) W_1 \\
X_2 &=& b + 1(\theta = 2) W_2 \\
X_3 &=& a + b + 1(\theta = 3) W_3
\end{eqnarray}
where $1(x)$ is the indicator function that takes value 1 if the event $x$ is true and 0 otherwise. Each message is assumed to be 1 independent uniform bit as well. 

\bigskip
\begin{figure}[h]
\begin{center}
\includegraphics[width= 5 in]{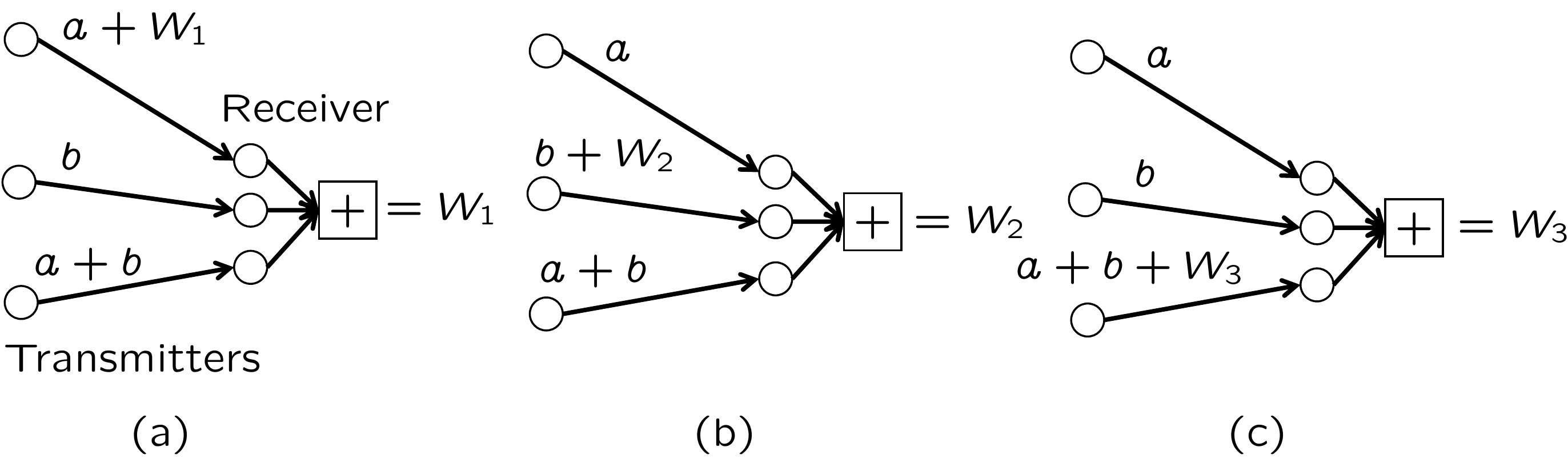}
\caption{\small The anonymous coding scheme with $K = 3$ transmitters. (a). $\theta = 1$. (b). $\theta = 2$. (c). $\theta = 3$. Note that no matter which message is sent, the receiver sees 3 uniform random bits and the decoding rule is always an addition. 
}
\label{fig:3tx}
\end{center}
\end{figure}

Correctness is easy to see as for all cases, the randomness cancels with each other after the addition operation. Anonymity holds because regardless of the value of $\theta$, the received signal consists of 3 uniform random bits and the decoding mapping is always an addition. As such, the receiver learns nothing about which transmitter is the source of the message.
We see that in order to communicate 1 bit anonymously, each transmitter needs to send 1 bit out. It is not hard to see that this is information theoretically optimal as even if there is no anonymity constraint, each transmitter will send out the desired message bit. What is non-trivial is the requirement on the correlated randomness. In this context, we show that for all linear schemes, 
each transmitter must hold a correlated random variable whose size is at least the size of the message and the total amount of randomness available at all transmitters must be at least as large as the size of $K-1$ messages. 
Further, when the scheme is capacity achieving, both the individual and total randomness sizes are optimal information theoretically (i.e., for all non-linear schemes as well).
A scheme of similar nature appears in a different context in \cite{Chaum_DC, Dolev_Ostrobsky}, where coded randomness is not allowed and optimality on the communications and randomness is not considered.

{\it Notation: 
For  integers $Z_1, Z_2, Z_1 \leq Z_2$, we use the compact notation $[Z_1:Z_2]=\{Z_1, Z_1+ 1,\cdots, Z_2\}$. 
The notation $X \sim Y$ is used to indicate that random variables $X$ and $Y$ are identically distributed. 
}

\section{Problem Statement}\label{sec:model}
Consider a network with $K$ transmitters and 1 receiver. Each transmitter is connected to the receiver with an orthogonal noiseless link. Each link can carry one symbol from a finite field $\mathbb{F}_p$ per channel use for a prime $p$.

Transmitter $k, k \in [1:K]$ has a message $W_k$. The messages $W_1, \cdots, W_K$ are independent and are each comprised of $L$ i.i.d. uniform symbols from $\mathbb{F}_p$. In $p$-ary units,
\begin{eqnarray}
H(W_{1}) &=& \cdots = H(W_{K}) = L, \label{h1}\\
H(W_{1}, \cdots, W_{K}) &=& H(W_{1}) + \cdots + H(W_{K}). \label{h2}
\end{eqnarray}

The transmitters wish to communicate with the receiver anonymously. The transmitters privately generate $\theta$ uniformly over $[1:K]$ (without loss of generality) and wish to communicate $W_\theta$ to the receiver while keeping $\theta$ a secret to the receiver. Depending on $\theta$, there are $K$ strategies that the transmitters employ to privately communicate the desired message\footnote{It turns out that for 
our achievable scheme, 
the transmitters do not need to know the exact value of the desired transmitter index $\theta$. It suffices for each transmitter to know that whether he is the desired or not.}. For example, if $\theta = k$, then in order to communicate $W_k$, Transmitter $i$ sends a signal $X_i^{[k]}$ over $N$ channel uses. To fulfill the task of communicating anonymously, we assume that Transmitter $i$ holds a correlated random variable $Z_i$. The correlated random variables are generated offline, i.e., before the realizations of the messages are known, so that the correlated random variables are independent of the messages.
\begin{eqnarray}
&&H(Z_1, \cdots, Z_K, W_1, \cdots, W_K) = 
H(Z_1, \cdots, Z_K) + H(W_1, \cdots, W_K) \label{zk_ind}
\end{eqnarray}
The transmitted signal, $X_i^{[k]}$, is a function of the information available at the transmitter (i.e., the message and the correlated random variable),
\begin{eqnarray}
H(X_i^{[k]} | W_i, Z_i) = 0 \label{det}
\end{eqnarray}
The received signal at the receiver is a collection of the $K$ transmitted signals.
\begin{eqnarray}
Y^{[k]} = [X_1^{[k]}, \cdots, X_K^{[k]}]^T \label{rec}
\end{eqnarray}

From $Y^{[k]}$, the receiver decodes the desired message $W_k$ according to a decoding mapping $g$. Note that the receiver is not allowed to learn anything about the index of the desired transmitter, so the decoding rule does not depend on $k$. The decoding mapping $g$ is fixed and known at every node (including the transmitters)\footnote{The encoding and decoding functions are globally known (akin to codebooks). Note that the receiver might try arbitrary other operations, for which no guarantees are made (e.g., correctness). Further note that the received signals are identically distributed, so the operations by the receiver do not depend on the desired message index (revealing no information).}.
\begin{eqnarray}
W_k = g(Y^{[k]}) \label{dec}
\end{eqnarray} 

To ensure transmitter anonymity, the $K$ strategies must be indistinguishable (identically distributed) from the perspective of the receiver, i.e., the following anonymity constraint must be satisfied $\forall k\in[1:K]$,
\begin{eqnarray}
\mbox{[Anonymity]}  ~~~~~~
(Y^{[1]}, g) &\sim& (Y^{[k]}, g) \notag\\
\mbox{i.e.,} ~~(X_1^{[1]}, \cdots, X_K^{[1]}, g) &\sim& (X_1^{[k]}, \cdots, X_K^{[k]}, g) \label{privacy}
\label{privacy}
\end{eqnarray}

The anonymous communication \emph{rate} characterizes how many symbols of desired information are communicated per symbol of total communication, and is defined as 
\begin{eqnarray}
R \triangleq \frac{L}{KN}
\end{eqnarray}
Note that by symmetry\footnote{Given any (asymmetric) achievable scheme that might employ a different number of channel uses for each transmitter, a symmetric scheme with the same rate (defined as the message size over the total number of channel uses by all transmitters) is obtained by repeating the original scheme $K$ times, and in the $i$-th repetition shifting the transmitter indices cyclicly by $i$.}, the number of channel uses for each transmitter does not depend on the transmitter indices.
A rate $R$ is said to be achievable if there exists an anonymous communication scheme of rate greater than or equal to $R$, for which zero error decoding is guaranteed.
The supremum of achievable rates is called the capacity $C$. 

The individual randomness size $\rho$ measures the amount of correlated randomness at each transmitter relative to the message size (by symmetry, without loss of generality, we assume that each transmitter holds the same amount of correlated randomness, i.e., $H(Z_1) = \cdots = H(Z_K)$). The total randomness size $\eta$ measures the total amount of correlated randomness at all transmitters relative to the message size.
\begin{eqnarray}
\rho &=& \frac{H(Z_1)}{L} \label{rho_def}\\
\eta &=& \frac{H(Z_1, \cdots, Z_K)}{L} \label{eta_def}
\end{eqnarray}

\section{Capacity of Anonymous Communications}
Theorem \ref{thm} states our main result.
\begin{theorem}\label{thm}
The capacity of anonymous communications over a parallel channel with $K$ transmitters and 1 receiver is 
$C = 1/K$. To achieve capacity, the minimum requirement on randomness size is $\rho = 1$ individually and $\eta = K-1$ in total.
\end{theorem}

The achievability proof appears in Section \ref{sec:ach}, where we provide a scalar linear anonymous coding scheme. The converse proof on the rate appears in Section \ref{sec:con}. 
The converse proof on the randomness appears in Section \ref{sec:con1} for linear schemes and Section \ref{sec:con2} for all possible schemes (i.e., the information theoretic converse).

When there is no anonymity constraint, the capacity is trivially 1 (only the desired transmitter sends its message) and no common randomness is needed. Therefore, in order to obtain anonymity among a set of $K$ transmitters, the price for anonymity in communication cost is $K$ times of that with no anonymity constraint and we further need $K-1$ bits of common randomness overall and 1 bit per transmitter, to communicate 1 bit anonymously.

\section{Proof of Theorem \ref{thm}: Achievabiliy}\label{sec:ach}
The achievable scheme with $K$ transmitters is an immediate generalization of that when $K = 3$, presented in the introduction section. We show that to communicate 1 bit anonymously, each transmitter uses its channel once, so that the rate achieved is $1/K$. 

We present the scheme over the binary field (any field will work in general).
Denote $a_1, \cdots, a_{K-1}$ as $K-1$ i.i.d. uniform bits, that are independent of the messages. The correlated random variables are assigned as follows.
\begin{eqnarray}
\begin{array}{l}
Z_i = a_i, ~i \in [1:K-1]  \\
Z_{K} = a_1 + \cdots + a_{K-1}
\end{array}
\end{eqnarray}
The transmitted signals are 
\begin{eqnarray}
\begin{array}{l}
X_i = a_i + 1(\theta = i) W_i,  ~i \in [1:K-1]\\
X_K =  a_1 + \cdots + a_{K-1} + 1(\theta = K) W_K \\
\end{array}
\end{eqnarray}
from which we can easily identify $X_i^{[k]}, \forall i, k \in [1:K]$.

The decoding mapping is the addition operation.
\begin{eqnarray}
g(Y) &=& X_1 + X_2 + \cdots + X_K \\
\mbox{i.e.,}~~g(Y^{[k]}) &=& X_1^{[k]} + X_2^{[k]} + \cdots + X_K^{[k]} = W_k
\end{eqnarray}
Correctness is easy to verify as the $K$ correlated random variables lie in a $K-1$ dimensional space (in fact, any $K-1$ dimensional space will work) and the decoding mapping is along the null space of the correlated random variables. Anonymity is guaranteed because for all possible values of $\theta$, the received signal is comprised of $K$ uniform i.i.d. bits and the decoding mapping does not depend on $\theta$. That is, when $\theta = k$, $\forall k \in [1:K]:$
\begin{eqnarray}
H(Y^{[k]}) &=& H(X_1^{[k]}, \cdots, X_K^{[k]}) \\
&=& H(a_1, a_2, \cdots, a_{K-1}, W_k) \\
&=& K
\end{eqnarray}

{\it Remark (Coded Randomness): In our coding scheme, the common randomness variables are correlated in coded form at the transmitters. Combining with the converse, we know that coded randomness is necessary to minimize the randomness size (i.e., if we do not allow randomness to be mixed, then we must use more randomness).}

{\it Remark (Collusion): Our achievable scheme is resilient to user collusions (equivalently, prior knowledge to preclude a set of non-desired transmitters) in the following sense. Suppose each transmitter only knows he is desired or not, then any collusion of $K-2$ non-desired transmitters with the receiver can not identify the desired transmitter index (i.e., the transmitters that are not in the colluding set are equally likely to be the desired).}

{\it Remark (Security): Our achievable scheme is perfectly secure in that the receiver obtains absolutely no information about all other messages beyond the desired one.}

\section{Proof of Theorem \ref{thm}: Converse on Rate}\label{sec:con}
We show that to transmit $L$ symbols anonymously, each transmitter must use the channel at least $N \geq L$ times. Then the rate bound $R = \frac{L}{NK} \leq 1/K
$ follows. 

We first show that $H(X_i^{[i]}) \geq L$, i.e., when Transmitter $i$ is the desired transmitter, he must send a signal that contains at least as much information as that contained in his message, from the correctness constraint. Define $W_{\bar{i}} = (W_1, \cdots, W_{i-1}, W_{i+1}, \cdots, W_K)$.
\begin{eqnarray}
L &\overset{(\ref{h1})}{=}& H(W_i) \\
&\overset{(\ref{dec})}{=}& I(W_i; Y^{[i]}) \\
&\overset{(\ref{rec})}{\leq}& I(W_i; X_1^{[i]}, \cdots, X_K^{[i]}, Z_1, \cdots, Z_K, W_{\bar{i}}) \\
&\overset{(\ref{zk_ind})(\ref{h2})}{=}& I(W_i; X_1^{[i]}, \cdots, X_K^{[i]} | Z_1, \cdots, Z_K, W_{\bar{i}}) \\
&\overset{(\ref{det})}{=}& I(W_i; X_i^{[i]} | Z_1, \cdots, Z_K, W_{\bar{i}}) \\
&\leq& H(X_i^{[i]}) \label{eq:desired}
\end{eqnarray}
Next, we show that  $H(X_i^{[k]}) \geq L, k \neq i$, i.e., when Transmitter $i$ is not the desired transmitter, he must send a statistically equivalent signal so that the entropy is also not less than the message size, from the anonymity constraint.
\begin{eqnarray}
H(X_i^{[k]}) &\overset{(\ref{privacy})}{=}& H(X_i^{[i]}) \\
&\overset{(\ref{eq:desired})}{\geq}& L, ~ k\neq i \label{eq:bl}
\end{eqnarray}

Combining with the fact that $H(X_i^{[k]}) \leq N, \forall k$, we arrive at the desired rate bound.

\section{
Proof of Theorem \ref{thm}: 
Converse on Randomness
for
Linear Schemes}\label{sec:con1}
We present the proof separately for linear schemes and all possible schemes (non-linear schemes included), because our result for linear schemes is stronger. 
We show that unconditionally, the individual randomness size $\rho \geq 1$ and sum randomness size $\eta \geq K-1$ for all linear schemes (with arbitrary positive rate). Otherwise, anonymous communication is not feasible, i.e., the capacity is 0.

\subsection{Proof for scalar linear case when $K=3$}
To illustrate the main idea in a simpler setting, we first consider the $K=3$ setting and assume the scheme is scalar linear, i.e., each message and each correlated random variable is only 1 symbol. We show that each correlated random symbol must be uniformly random, $H(Z_i) \geq L, i \in \{1,2,3\}$ and any two random symbols are independent, $H(Z_i, Z_j) \geq 2L, i \neq j, i, j \in \{1,2,3\}$.

For a linear scheme, the transmitted signal is a linear combination of the message symbol and the correlated random variable, and the decoding mapping is also a linear combination of the received signal symbols (so the only operation allowed is taking linear combinations). Specifically, the transmitted signals are
\begin{eqnarray}
X_i^{[k]} &=& V_i^{[k]} W_i + U_i^{[k]} Z_i, ~~i,k \in \{1,2,3\} \label{eq:xdef}
\end{eqnarray}
where $V_i^{[k]}, U_i^{[k]}$ are scalars over $\mathbb{F}_p$.
The decoding coefficients are denoted as $G_1, G_2, G_3 \in \mathbb{F}_p$ (note that the constants $G_1, G_2, G_3$ do not depend on the desired transmitter index $k$) and the decoding works as follows.
\begin{eqnarray}
W_k &=& G_1 X_1^{[k]} + G_2 X_2^{[k]} + G_3 X_3^{[k]} \\
&=& G_1 V_1^{[k]} W_1 + G_2 V_2^{[k]} W_2 + G_3 V_3^{[k]} W_3 
+ G_1 U_1^{[k]} Z_1 + G_2 U_2^{[k]} Z_2 + G_3 U_3^{[k]} Z_3 \label{eq:wk}
\end{eqnarray}
As such, for any $k \in \{1,2,3\}$, the undesired messages can not appear. It follows from the equality (\ref{eq:wk}) that
\begin{eqnarray}
&& G_1 V_1^{[1]}  \neq 0, G_2 V_2^{[1]} = 0, G_3 V_3^{[1]} = 0 \\
&& G_1 V_1^{[2]}  = 0, G_2 V_2^{[2]} \neq 0, G_3 V_3^{[2]} = 0 \\
&& G_1 V_1^{[3]}  = 0, G_2 V_2^{[3]} = 0, G_3 V_3^{[3]} \neq 0 \\
&\Rightarrow& G_1 \neq 0, G_2 \neq 0, G_3 \neq 0, 
V_1^{[1]} \neq 0, V_2^{[2]} \neq 0, V_3^{[3]}\neq 0, \notag\\
&& V_2^{[1]} = V_3^{[1]} = 0, V_1^{[2]} = V_3^{[2]} = 0, V_1^{[3]} = V_3^{[3]} = 0 
\label{eq:v1}
\end{eqnarray}
Consider now $X_1^{[2]} \overset{}{=} U_1^{[2]} Z_1$. From (\ref{eq:bl}), we have
\begin{eqnarray}
L &\overset{(\ref{eq:bl})}{\leq}& H(X_1^{[2]}) \\
&\overset{(\ref{eq:xdef})(\ref{eq:v1})}{=}& H(U_1^{[2]} Z_1)  \\
&=& H(Z_1) \label{eq:notzero} \\
&\overset{(\ref{rho_def})}{=}& \rho L
\end{eqnarray}
where (\ref{eq:notzero}) follows from the observation that 
\begin{eqnarray}
U_1^{[2]} \neq 0, \label{eq:u12}
\end{eqnarray} 
as otherwise $H(X_1^{[2]}) = 0$, contradicting (\ref{eq:bl}). Therefore, we have proved that the individual randomness size $\rho \geq 1$. 
Symmetrically, from (\ref{eq:notzero}) and (\ref{eq:u12}), we have
\begin{eqnarray}
&& L \leq H(Z_2), L \leq H(Z_3), \label{eq:z2}\\
&& U_i^{[k]} \neq 0, ~k \neq i \label{eq:uik}
\end{eqnarray}
Next, we consider $(X_1^{[1]}, X_2^{[1]}) = (V_1^{[1]} W_1 + U_1^{[1]}Z_1, U_2^{[1]} Z_2)$.
\begin{eqnarray}
&& H(X_1^{[1]}, X_2^{[1]}) \notag \\
&\overset{(\ref{eq:xdef})(\ref{eq:v1})}{=}& H(U_2^{[1]} Z_2) + H(V_1^{[1]} W_1 + U_1^{[1]}Z_1 | U_2^{[1]} Z_2) 
\\
&\overset{(\ref{eq:uik})}{\geq}& H(Z_2) + H(V_1^{[1]} W_1 + U_1^{[1]}Z_1 | Z_2, Z_1) \\
&\overset{(\ref{eq:z2})}{\geq}& L + H(V_1^{[1]} W_1 | Z_2, Z_1) \\
&\overset{(\ref{zk_ind})}{=}& L + H(V_1^{[1]} W_1) \\
&\overset{(\ref{eq:v1})(\ref{h1})}{=}& 2L \label{eq:l2}
\end{eqnarray}
Then we consider $H(X_1^{[3]}, X_2^{[3]}) \overset{(\ref{eq:xdef})(\ref{eq:v1})}{=} H(U_1^{[3]} Z_1, U_2^{[3]} Z_2)$, as follows.
\begin{eqnarray}
\eta L &\overset{(\ref{eta_def})}{=}& H(Z_1, Z_2, Z_3) \\
&\geq& H(Z_1, Z_2) \\
&\overset{(\ref{eq:xdef})(\ref{eq:v1})(\ref{eq:uik})}{=}& H(X_1^{[3]}, X_2^{[3]}) \\
&\overset{(\ref{privacy})}{=}& H(X_1^{[1]}, X_2^{[1]}) \\
&\overset{(\ref{eq:l2})}{\geq}& 2L
\end{eqnarray}
Therefore we have proved that the sum randomness size $\eta \geq 2 = K-1$.

{\it Remark: From (\ref{eq:wk}), we know that the correlated random variables must satisfy some linear equation, i.e., they must lie in a lower dimensional space (rank deficient) for successful decoding.}

\subsection{General proof for vector linear case with arbitrary $K$} 
We generalize the above proof to the vector linear case with arbitrary number of transmitters, $K$. We show that $H(Z_1) \geq L$ and $H(Z_1, \cdots, Z_{K-1}) \geq (K-1)L$. 

The vector linear scheme is represented as follows.
\begin{eqnarray}
X_i^{[k]} &=& {\bf V}_i^{[k]} W_i + {\bf U}_i^{[k]} Z_i, ~~i,k \in [1:K] \label{eq:xdefk}
\end{eqnarray}
where ${\bf V}_i^{[k]}, {\bf U}_i^{[k]}$ are $N \times L$ constant encoding matrices, over $\mathbb{F}_p$ (and are globally known). Note that there is no loss of generality in assuming that $Z_i$ contains $L$ symbols over $\mathbb{F}_p$, as we do not impose any statistical properties on the $L$ symbols (e.g., they are not necessarily independent).
For any $i, k \in [1:K]$,
\begin{eqnarray}
W_k &=& \sum_{i=1}^K {\bf G}_i X_i^{[k]} 
\\  
&=& \sum_{i=1}^K {\bf G}_i {\bf V}_i^{[k]} W_i + \sum_{i=1}^K {\bf G}_i {\bf U}_i^{[k]} Z_i
\end{eqnarray}
The decoding mapping is specified by the constant filtering matrices ${\bf G}_i$, which have dimension $L \times N$ over $\mathbb{F}_p$. Then we have
\begin{eqnarray}
&& \mbox{rank}({\bf G}_k {\bf V}_k^{[k]}) = L, ~~ k \in [1:K] \label{eq:fr} \\
&& {\bf G}_k {\bf V}_k^{[i]} = 0, ~~ k \neq i, ~i,k \in [1:K] \label{eq:nc}
\end{eqnarray}
Following the proof presented in the previous section, we proceed to consider ${\bf G}_1 X_1^{[2]} \overset{(\ref{eq:xdefk})(\ref{eq:nc})}{=} {\bf G}_1 {\bf U}_1^{[2]} Z_1$.
\begin{eqnarray}
L &\overset{(\ref{h1})}{=}&  H(W_1) \\
&\overset{(\ref{eq:fr})}{=}& H({\bf G}_1 {\bf V}_1^{[1]} W_1) \\
&\overset{(\ref{zk_ind})}{=}& H({\bf G}_1 {\bf V}_1^{[1]} W_1 | Z_1)  \\
&\overset{(\ref{eq:xdefk})}{=}& H({\bf G}_1 X_1^{[1]} | Z_1)\\
&\leq& H({\bf G}_1 X_1^{[1]}) \\
&\overset{(\ref{privacy})}{=}& H({\bf G}_1 X_1^{[2]})  \label{eq:ind1} \\
&\overset{(\ref{eq:xdefk})(\ref{eq:nc})}{=}& H({\bf G}_1 {\bf U}_1^{[2]} Z_1) \label{eq:vz}\\
&\leq& H(Z_1) \\
&\overset{(\ref{rho_def})}{=}& \rho L
\end{eqnarray}
Therefore, we have proved that the individual randomness size $\rho \geq 1$. 
As a byproduct, from (\ref{eq:vz}), we obtain that
\begin{eqnarray}
\mbox{rank}({\bf G}_1 {\bf U}_1^{[2]}) = L \label{eq:uik1}
\end{eqnarray} 
as otherwise we have the contradiction that $H({\bf G}_1 {\bf U}_1^{[2]} Z_1) < L$.
Symmetrically, from (\ref{eq:uik1}), we have
\begin{eqnarray}
&& \mbox{rank}({\bf G}_k {\bf U}_k^{[i]}) = L, ~k \neq i \label{eq:uikg}
\end{eqnarray}
Next, we consider the total randomness size. We first prove a lemma.
\begin{lemma}\label{lemma:full}
For all $ i \in [1:K-1]$, we have
\begin{eqnarray}
H({\bf G}_1 X_1^{[i+1]}, {\bf G}_2 X_2^{[i+1]}, \cdots, {\bf G}_{i} X_{i}^{[i+1]}) \geq i L \label{eq:lemma}
\end{eqnarray}
\end{lemma}
{\it Proof:} The proof is based on induction. Note that the basis case where $i = 1$ is proved in (\ref{eq:ind1}). Suppose now (\ref{eq:lemma}) holds when $i = j, j \in [1:K-2]$, i.e.,
\begin{eqnarray}
H({\bf G}_1 X_1^{[j+1]}, {\bf G}_2 X_2^{[j+1]}, \cdots, {\bf G}_{j} X_{j}^{[j+1]}) \geq jL \label{eq:ind2}
\end{eqnarray}
Now consider the case where $i = j+1$.
\begin{eqnarray}
&& H({\bf G}_1 X_1^{[j+2]}, {\bf G}_2 X_2^{[j+2]}, \cdots, {\bf G}_{j+1} X_{j+1}^{[j+2]}) \notag \\
&\overset{(\ref{privacy})}{=}& H({\bf G}_1 X_1^{[j+1]}, {\bf G}_2 X_2^{[j+1]}, \cdots, {\bf G}_{j+1} X_{j+1}^{[j+1]})  
\\
&=& H({\bf G}_1 X_1^{[j+1]}, {\bf G}_2 X_2^{[j+1]}, \cdots, {\bf G}_{j} X_{j}^{[j+1]}) \notag\\
&&+~ H({\bf G}_{j+1} X_{j+1}^{[j+1]} | {\bf G}_1 X_1^{[j+1]}, \cdots, {\bf G}_{j} X_{j}^{[j+1]}) 
\\
&\overset{(\ref{eq:ind2})(\ref{eq:xdefk})(\ref{eq:nc})}{\geq}& jL + H( {\bf G}_{j+1} X_{j+1}^{[j+1]} | Z_1, \cdots, Z_{j}, Z_{j+1}) \\
&\overset{(\ref{eq:xdefk})(\ref{eq:nc})}{=}& jL + H( {\bf G}_{j+1} {\bf V}_{j+1}^{[j+1]} W_{j+1} | Z_1, \cdots, Z_{j+1}) 
\\
&\overset{(\ref{zk_ind})(\ref{eq:fr})}{=}& jL + H(W_{j+1}) \\
&\overset{(\ref{h1})}{=}& (j+1)L \label{eq:l2k}
\end{eqnarray}
Since both the basis and the inductive steps have been performed, by mathematical induction, we have proved that (\ref{eq:lemma}) holds for all $i \in [1:K-1]$. The proof for Lemma \ref{lemma:full} is complete.

\hfill\QED

Finally, consider (\ref{eq:lemma}) and set $i = K-1$. We have
\begin{eqnarray}
(K-1)L &\overset{(\ref{eq:lemma})}{\leq}& H(X_1^{[K]}, \cdots, X_{K-1}^{[K]}) \\
&\overset{(\ref{eq:xdefk})(\ref{eq:nc})(\ref{eq:uikg})}{\leq}& H(Z_1, \cdots, Z_{K-1}) \\
&\overset{(\ref{eta_def})}{\leq}& \eta L 
\end{eqnarray}
Therefore we have proved that the sum randomness size $\eta \geq K-1$, for any rate $R = \frac{L}{NK}$.

\section{
Proof of Theorem \ref{thm}: 
Information Theoretic 
Converse on Randomness for 
Capacity Achieving Schemes} \label{sec:con2}
We show that when the scheme is capacity achieving, i.e., the rate achieved is $1/K$, i.e., $H(X_i^{[k]}) = N = L, \forall i,k \in [1:K]$, then the randomness sizes $\rho = 1$ and $\eta = K-1$ are both information theoretically optimal. 
\subsection{Proof for 
binary scalar case when $K = 3$}
Before presenting the general proof for arbitrary $K$, we first consider the $K =3$ case and assume that each message is one bit, to illustrate the idea. Then in this case, $L = 1$ and the field is $\mathbb{F}_2$. In this case, we need to show that $H(Z_i) \geq 1$ and $H(Z_1, Z_2, Z_3) \geq 2$.

First, for capacity achieving schemes, i.e., 
\begin{eqnarray}
H(X_i^{[k]}) = 1, \forall i,k \in \{1,2, 3\} \label{eq:el}
\end{eqnarray} 
the received signal is uniformly random. The proof is deferred to Lemma \ref{lemma:uniform} for the general case. That is, for any $k$,
\begin{eqnarray}
X_1^{[k]}, X_2^{[k]}, X_3^{[k]} ~\mbox{is uniformly distributed.} 
\label{eq:samedis}
\end{eqnarray}

Next, consider $X_1^{[2]}, X_2^{[2]}, X_3^{[2]}, W_2$. Note that
\begin{eqnarray}
&& H(W_2 | X_1^{[2]}, X_3^{[2]}) \overset{(\ref{h2})(\ref{zk_ind})(\ref{det})}{=} H(W_2) \overset{(\ref{h1})}{=} L \\
&& H(X_2^{[2]}| X_1^{[2]}, X_3^{[2]}) \overset{(\ref{eq:samedis})}{=} L \\
&& H(W_2 | X_1^{[2]}, X_2^{[2]}, X_3^{[2]}) \overset{(\ref{dec})}{=} 0 
\end{eqnarray}
Then we have the observation that for any realization of $X_1^{[2]}, X_3^{[2]}$, $W_2$ has a one-to-one mapping to $X_2^{[2]}$, i.e.,
\begin{eqnarray}
H(X_2^{[2]} | W_2, X_1^{[2]}, X_3^{[2]}) = 0 \label{eq:e0}
\end{eqnarray}

Repeating the argument for $W_1$ and $W_3$, we have
\begin{eqnarray}
&& H(X_1^{[1]} | W_1, X_2^{[1]}, X_3^{[1]}) = 0 \label{eq:e1} \\
&& H(X_3^{[3]} | W_3, X_1^{[3]}, X_2^{[3]}) = 0 \label{eq:e3}
\end{eqnarray}

From the anonymity constraint (\ref{privacy}) and the correctness constraint (\ref{dec}), we know that
\begin{eqnarray}
(X_1^{[1]}, X_2^{[1]}, X_3^{[1]}, g, W_1) \sim (X_1^{[k]}, X_2^{[k]}, X_3^{[k]}, g, W_k) \label{eq:sprivacy}
\end{eqnarray}

We now consider the individual randomness size. Combining (\ref{eq:e0}) and (\ref{eq:sprivacy}), we have
\begin{eqnarray}
H(X_2^{[1]} | W_1, X_1^{[1]}, X_3^{[1]}) = 0 \label{eq:x2det}
\end{eqnarray}
Then
\begin{eqnarray}
&& I(X_2^{[1]} ; W_2) \notag\\
&\leq& I(X_2^{[1]}, W_1, X_1^{[1]}, X_3^{[1]}; W_2) \\
&=& I(W_1, X_1^{[1]}, X_3^{[1]}; W_2) 
+ I(X_2^{[1]}; W_2 | W_1, X_1^{[1]}, X_3^{[1]}) \\
&\overset{(\ref{eq:x2det})}{\leq}& I(W_1, X_1^{[1]}, Z_1, X_3^{[1]}, W_3, Z_3; W_2) + 0 \\
&\overset{(\ref{h2})(\ref{zk_ind})(\ref{det})}{=}& 0 \label{eq:x2w2}
\end{eqnarray}
and
\begin{eqnarray}
1 &\overset{(\ref{eq:el})}{=}& H(X_2^{[1]}) \\
&\overset{(\ref{det})}{=}& I(X_2^{[1]}; W_2, Z_2) \\
&\overset{(\ref{eq:x2w2})}{=}& I(X_2^{[1]}; Z_2 |W_2) \\
&\leq& H(Z_2) \\
&\overset{(\ref{rho_def})}{=}& \rho 
\end{eqnarray}
Therefore the individual randomness size satisfies that $\rho \geq 1$.

We proceed next to consider the sum randomness size. Combining (\ref{eq:e0}), (\ref{eq:e1}), (\ref{eq:e3}) and (\ref{eq:sprivacy}), we have obtained the structure of the decoding mapping, i.e., for any $3$-tuple of the received signal, if 2 elements are fixed, the remaining element has a one-to-one mapping with the desired message. For example, when $Y^{[k]} = (0,0,0)$, suppose that $W_k = g(Y^{[k]}) = g(0,0,0) = w, w \in \{0,1\}$, then $g(0,0,1) = g(0,1,0) = g(1,0,0) = 1 - w$. Proceeding along this line, the decoding mapping is uniquely identified as follows.
\begin{eqnarray}
\begin{array}{c|c}
Y^{[k]} & W_k \\ \hline 
(0,0,0) & w \\
(0,0,1) & 1-w \\
(0,1,0) & 1-w \\
(0,1,1) & w \\
(1,0,0) & 1-w \\
(1,0,1) & w \\
(1,1,0) & w \\
(1,1,1) & 1-w
\end{array}
\end{eqnarray}
We are now ready to show that 
\begin{eqnarray}
H(X_2^{[1]}, X_3^{[1]} | Z_1, Z_2, Z_3) = 0. \label{eq:randomdet}
\end{eqnarray}
Consider an arbitrary realization of $(W_1, Z_1, Z_2, Z_3) = (w_1, z_1, z_2, z_3)$, drawn according to the correct joint distribution ($W_1$ is independent of $Z_1, Z_2, Z_3$). Then $X_1^{[1]}$ is a constant (denoted as $x_1$) as $X_1^{[1]}$ is a function of $W_1$ and $Z_1$. We now show that $X_2^{[1]}, X_3^{[1]}$ are now constants as well. Note that the only variables that are random now are $W_2, W_3$. Suppose $X_2^{[1]}$ is still random, depending on the value of $W_2$. Then consider two realizations of $X_2^{[1]}$ (denoted as $x_2, x_2', x_2 \neq x_2'$) and the received signal realizations
\begin{eqnarray}
y_1 = (x_1, x_2, X_3^{[1]}) \\
y_2 = (x_1, x_2', X_3^{[1]})
\end{eqnarray}
From the decoding mapping table, we know that $g(y_1) \neq g(y_2)$. However, from the correctness constraint, we know that $g(y_1) = g(y_2) = w_1$. Therefore, we arrive at the contradiction and $X_2^{[1]}, X_3^{[1]}$ are deterministic functions of the correlated random variables. Then we have
\begin{eqnarray}
\eta &\overset{(\ref{eta_def})}{=}& H(Z_1, Z_2, Z_3) \\
&\overset{(\ref{eq:randomdet})}{=}& H(X_2^{[1]}, X_3^{[1]}, Z_1, Z_2, Z_3) \\
&\geq& H(X_2^{[1]}, X_3^{[1]}) \\
&\overset{(\ref{eq:samedis})}{=}& 2
\end{eqnarray}
Therefore the sum randomness size $\eta \geq 2$ and the proof is complete.

\subsection{Proof for 
Arbitrary $K$}
We follow the steps of the proof for $K=3$ binary case and show $H(Z_i) \geq L, H(Z_1, Z_2, \cdots, Z_K) \geq (K-1)L$.

First, we present a lemma, which says that the received signals are uniformly random, when the scheme is capacity achieving.
\begin{lemma}\label{lemma:uniform}
\begin{eqnarray}
&& H(X_i^{[k]}) = N = L, \forall i,k \in [1:K] \label{eq:elk} \\
&\Rightarrow&  H(X_1^{[k]}, \cdots, X_K^{[k]}) = KL, \forall k \in [1:K]  \label{eq:samedisg}
\end{eqnarray}
\end{lemma}
{\it Proof:}
Note that (\ref{eq:elk}) implies that $H(X_1^{[k]}, \cdots, X_K^{[k]}) \leq KL$. It suffices to prove only the other direction. Define $X_{\bar{i}}^{[i]} = (X_1^{[i]}, \cdots. X_{i-1}^{[i]}, X_{i+1}^{[i]}, \cdots, X_K^{[i]})$.
\begin{eqnarray}
&& H(X_1^{[k]}, \cdots, X_K^{[k]}) \notag\\
&\overset{}{=}& \sum_{i=1}^K H(X_i^{[k]} | X_1^{[k]}, \cdots, X_{i-1}^{[k]}) 
\\
&\overset{(\ref{privacy})}{=}& \sum_{i=1}^K H(X_i^{[i]} | X_1^{[i]}, \cdots, X_{i-1}^{[i]}) \\
&\geq& \sum_{i=1}^K H(X_i^{[i]} | X_{\bar{i}}^{[i]}) \\
&\geq& \sum_{i=1}^K I(W_i; X_i^{[i]} | X_{\bar{i}}^{[i]})  \\
&\overset{(\ref{dec})}{=}& \sum_{i=1}^K H(W_i | X_{\bar{i}}^{[i]})  \label{eq:w2k}\\
&\overset{}{\geq}& \sum_{i=1}^K H(W_i | X_{\bar{i}}^{[i]}, W_{\bar{i}}, Z_1, \cdots, Z_K)  \\
&\overset{(\ref{det})(\ref{zk_ind})(\ref{h1})}{=}& KL
\end{eqnarray}

\hfill\QED

Next, note that
\begin{eqnarray}
&& H(W_i | X_{\bar{i}}^{[i]}) \overset{(\ref{eq:w2k})}{=} L, \\
&& H(X_{i}^{[i]} | X_{\bar{i}}^{[i]}) \overset{(\ref{eq:samedisg})}{=} L, \\
&& H(W_i | X_{\bar{i}}^{[i]}, X_{i}^{[i]}) \overset{(\ref{dec})}{=} 0 
\end{eqnarray}
Note that $W_i$ is independent of $X_{\bar{i}}^{[i]}$.
Then we have the observation that for any realization of $X_{\bar{i}}^{[i]}$, $W_i$ has a one-to-one mapping to $X_{i}^{[i]}$, i.e.,
\begin{eqnarray}
H(X_i^{[i]} | W_i, X_{\bar{i}}^{[i]}) = 0 \label{eq:e0k}
\end{eqnarray}
From the anonymity constraint (\ref{privacy}) and the correctness constraint (\ref{dec}), we know that for any $i \in [1:K]$,
\begin{eqnarray}
(X_i^{[i]}, X_{\bar{i}}^{[i]}, g, W_i) \sim (X_i^{[1]}, X_{\bar{i}}^{[1]}, g, W_1) \label{eq:sprivacyk}
\end{eqnarray}
Combining (\ref{eq:e0k}) and (\ref{eq:sprivacyk}), we have
\begin{eqnarray}
H(X_i^{[1]} | W_1, X_{\bar{i}}^{[1]}) = 0 \label{eq:x2detk}
\end{eqnarray}
Then for $i \neq 1$,
\begin{eqnarray}
&& I(X_i^{[1]} ; W_i) \notag\\
&\leq& I(X_i^{[1]}, W_1, X_{\bar{i}}^{[1]}; W_i) \\
&=& I(W_1, X_{\bar{i}}^{[1]}; W_i) + I(X_i^{[1]}; W_i | W_1, X_{\bar{i}}^{[1]}) \\
&\overset{(\ref{eq:x2detk})}{\leq}& I(W_1, X_{\bar{i}}^{[1]}, Z_{\bar{i}}, W_{\bar{i}}; W_i) + 0 \\
&\overset{(\ref{zk_ind})(\ref{det})}{=}& 0 \label{eq:x2w2k}
\end{eqnarray}
where $Z_{\bar{i}} = (Z_1, \cdots, Z_{i-1}, Z_{i+1}, \cdots, Z_K)$.

For the individual randomness size, we have
\begin{eqnarray}
L &\overset{(\ref{eq:elk})}{=}& H(X_i^{[1]}) \\
&\overset{(\ref{det})}{=}& I(X_i^{[1]}; W_i, Z_i) \\
&\overset{(\ref{eq:x2w2k})}{=}& I(X_i^{[1]}; Z_i |W_i) \\
&\leq& H(Z_i) \\
&\overset{(\ref{rho_def})}{=}& \rho L
\end{eqnarray}
Therefore $\rho \geq 1$. 

For the sum randomness size, as (\ref{eq:e0k}) holds for all $i \in [1:K]$ and from (\ref{eq:sprivacyk}), we know that if any $K-1$ elements of the received signal are determined, the remaining element has a one-to-one mapping with the desired message, which means that
\begin{eqnarray}
&&\mbox{For 2 received signal tuples that differ in 1 element,} \notag\\
&&\mbox{i.e.,}~ y_1 = (x_1, \cdots, x_k, \cdots, x_K), \notag\\
&& ~~~~~y_2 = (x_1, \cdots, x_k', \cdots, x_K), \notag\\
&& \mbox{we have}~ g(y_1) \neq g(y_2). \label{eq:diff}
\end{eqnarray}
Then we claim that $X_2^{[1]}, \cdots, X_K^{[1]}$ are functions of $Z_1, \cdots, Z_K$. This result is stated in the following lemma.
\begin{lemma}\label{lemma:det}
\begin{eqnarray}
H(X_2^{[1]}, \cdots, X_K^{[1]} | Z_1, \cdots, Z_K) = 0 \label{eq:lemma2}
\end{eqnarray}
\end{lemma}
{\it Proof:} Consider one arbitrary realization of $W_1, Z_1, \cdots, Z_K$, denoted as $(W_1, Z_1, \cdots, Z_K) = (w_1, z_1, \cdots, z_K)$. As $W_1, Z_1$ are fixed, then $X_1^{[1]}$ is a constant, denoted as $x_1$. We show that $X_2^{[1]}, \cdots, X_K^{[1]}$ are constants now. To set up the proof by contradiction, suppose there exists one $X_k^{[1]}$ that can take multiple values. Denote two such values as $x_k, x_k', x_k \neq x_k'$. The other $X_i^{[1]}, i \neq k$ are assumed to be constants and denoted as $x_i$. Note that for fixed $z_2. \cdots, z_k$, $X_2^{[1]}, \cdots, X_K^{[1]}$ are conditionally independent as now the randomness only comes from the messages $W_2, \cdots, W_K$ and the messages are independent. We now have two different received signal tuples
\begin{eqnarray}
y_1 = (x_1, \cdots, x_k, \cdots, x_K)\\
y_2 = (x_1, \cdots, x_k', \cdots, x_K)
\end{eqnarray}
From (\ref{eq:diff}), we know that $g(y_1) \neq g(y_2)$. However, this contradicts with the fact that $g(y_1) = g(y_2) = w_1$. Therefore we have arrived at the contradiction and $X_2^{[1]}, \cdots, X_K^{[1]}$ are functions of $Z_1, \cdots, Z_K, W_1$. Further $X_2^{[1]}, \cdots, X_K^{[1]}$ are independent of $W_1$ and we have proved the lemma.

\hfill\QED

From Lemma \ref{lemma:det}, we have
\begin{eqnarray}
\eta L &\overset{(\ref{eta_def})}{=}& H(Z_1, \cdots, Z_K) \\
&\overset{(\ref{eq:lemma2})}{=}& H(X_2^{[1]}, \cdots, X_K^{[1]}, Z_1, \cdots, Z_K) \\
&\geq& H(X_2^{[1]}, \cdots, X_K^{[1]}) \\
&\overset{(\ref{eq:elk})}{=}& (K-1) L
\end{eqnarray}
Therefore the desired sum randomness size bound follows and the proof is complete.

{\it Remark: The above proof relies on the assumption that the scheme is capacity achieving. Otherwise, Lemma \ref{lemma:uniform} and Lemma \ref{lemma:det} may not hold.}

{\color{black}
{\it Remark: The individual randomness size bound holds without the constraint that the achieved rate is equal to the capacity, i.e., we have $\rho \geq 1$ for any positive rate (the total randomness size bound, however, hinges on the assumption of capacity achieving schemes). A sketch of proof idea is as follows (the above proof is more informative in that the combinatoric structure of the decoding mapping is revealed). We first note that the transmitted signal from Transmitter $i, i\neq 1$ is independent of $W_1$, i.e., $I(X_i^{[1]} ; W_1) = 0$. Next, from the anonymity constraint, the same relation on the mutual information must hold when $W_i$ is desired, i.e., $I(X_i^{[i]} ; W_i) = 0$, meaning that the transmitted signal from Transmitter $i$ does not contain any information about $W_i$. To guarantee this, the randomness needed must be at least as large as the message size.
}

{\it Remark: A more general condition where the bound on the total randomness size holds unconditionally for arbitrary positive rates is when we require the transmitted signal to be deterministic functions of the correlated random variable when he is not desired, i.e., $H(X_i^{[k]} | Z_i) = 0, i \neq k$ (in other words, the messages do not play a role when they are not desired. As the messages are independent among themselves and of the correlated random variables, it will be interesting if they help to reduce total randomness size). Lemma \ref{lemma:det} proves that this deterministic condition holds for capacity achieving schemes. After we assume this deterministic condition to be satisfied, the proof is the same as that presented above after Lemma \ref{lemma:det}.
}

%
}

\section{Conclusion}
We consider the problem of anonymous communications from an information theory perspective. We have characterized the capacity of anonymous communications over a parallel channel with $K$ transmitters and $1$ receiver, to be $C = 1/K$. Further, the minimum randomness sizes required are $\rho = 1$ per transmitter and $\eta = K-1$ for all transmitters.
This work represents a step towards using information theoretic tools to understand the fundamental limits of anonymous network communications.

\let\url\nolinkurl
\bibliographystyle{IEEEtran}
\bibliography{Thesis}
\end{document}